# High-resolution optical frequency dissemination on a telecommunication network with data traffic

Fabien Kéfélian[1], Olivier Lopez[1], Haifeng Jiang[2], Christian Chardonnet[1], Anne Amy-Klein[1*], and Giorgio Santarelli[2]

[1] *Laboratoire de Physique des Lasers, CNRS, Université Paris 13, 99 av. J.-B. Clément, 93430 Villetaneuse, France*

[2] *LNE-SYRTE, Observatoire de Paris, CNRS, UPMC, 61 Avenue de l'Observatoire, 75014 Paris, France*

[*]*Corresponding author : amy@univ-paris13.fr*

We transferred the frequency of an ultra-stable laser over a 108 km urban fiber link comprising 22 km of optical communications network fiber simultaneously carrying Internet data traffic. The metrological signal and the digital data signal are transferred on two different frequency channels in a dense wavelength division multiplexing scheme. The metrological signal is inserted into and extracted from the communications network by using bidirectional off-the-shelf optical add-drop multiplexers. The link-induced phase noise is measured and cancelled with round-trip technique using an all-fiber-based interferometer. The compensated link shows an Allan deviation of a few $10^{-16}$ at one second and below $10^{-19}$ at 10,000 seconds. This opens the way to a wide dissemination of ultra stable optical clock signals between distant laboratories via the Internet network.

2009 Optical Society of America

The transfer of ultra-stable frequencies between distant laboratories is an important issue for a large number of high-sensitivity experiments in metrology and fundamental physics, for example tests of the fundamental constants stability [1] or high-resolution geodetic measurements. With the recent development of single ion traps and cold atoms lattices, optical clocks are expected to reach fractional frequency instability level of $10^{-17}$ or better at one day [2-3], which implies the reduction of transfer induced instability well below this level.

The transmission of frequency standards over optical fiber has been investigated for a few years [4-14]. Early optical fiber links use amplitude modulation of an optical carrier to transmit RF and microwave frequencies [4-7]. However to take full advantage of the optical fiber media the direct optical frequency transfer is the appropriate choice. Indeed, thanks to heterodyne detection the effect of link attenuation halves in dB, and, due to the higher carrier frequency the link-induced phase noise can be detected with much higher resolution. Since 2007, optical frequency transfer over a fiber link of more than 100 km has been reported by several groups [8-13] and has demonstrated the feasibility of a full optical link with instability in the $10^{-17}$ to $10^{-19}$ range. These experiments used dedicated dark fibers (i.e. not currently used for telecommunications) in urban environment. Since last year a long link of about 1000 km has connected several German research laboratories using dark fibers from the German national telecommunication network for research and education (DFN) and the stabilization of this link is currently under development [13]. These are the first milestones towards continental scale fiber links, but this approach requires dedicated fibers between every laboratory to be connected.

In this paper, we report a new approach for ultrastable frequency transfer which consists in using the already existing optical telecommunication network and transmitting the optical frequency reference together with the internet data traffic, as already realized for very low resolution frequency dissemination [14]. We demonstrate here the transmission of an optical frequency reference over 108 km by extending 86 km of dedicated fiber with 22 km of fiber from the French National Telecommunication network for Technology, Education and Research (RENATER) carrying simultaneously digital data. First, we present the scheme of the hybrid link and the interconnection method between the dark fiber span and the RENATER network. The frequency noise and the stability performance are then reported and discussed.

The metrological reference signal (RS) transferred through the link is provided by a sub-Hz-linewidth cavity-stabilized laser emitting at 1542,14 nm (ITU channel #44) [11]. Fig. 1 shows the overall scheme of the 108 km optical link, which starts and ends at Observatoire de Paris to characterize the link stability, and comprises four spans. The first span is a dedicated dark fiber from the urban network which connects two laboratories, LPL (at University Paris 13, in the north surroundings of Paris) and LNE-SYRTE (at Observatoire de Paris), and extends over 43 km [4]. The second and third spans are two 11 km-long fibers connecting the information services and technology center of Université Paris 13 to a node of the French National Telecommunications network for Technology, Education and Research (RENATER) located in Aubervilliers. The digital stream between Université Paris 13 and Aubervilliers node is encoded on an optical carrier on the ITU channel #34 (1550,12 nm) within these two upstream and downstream fibers using 10 GBit/s data stream technology. At Aubervilliers the RS arriving from one fiber is sent into the other 11 km fiber toward Université Paris 13 and then into a second 43 km dark fiber linking Université Paris 13 to Observatoire de Paris.

Interconnection between the 43 km dark fibers and the optical telecommunication network is not straightforward. Concerning RS transmission, the round-trip propagation of the optical signal is required on the same fiber for noise compensation. For the Internet signal transmission, it is necessary to minimize insertion losses and avoid cross-talk. An elegant solution is to use bidirectional off-the-shelf optical add-drop multiplexer (OADM). This three-port component can insert or extract a specified wavelength from the other wavelengths, with isolation better than 25 dB for an adjacent channel (100 GHz) and better than 40 dB for other channels. The losses are about 1.2 dB for the add/drop channel and below 1 dB for the other channels. Four OADMs are used along the link to insert and extract the RS on the upstream and downstream 11 km fibers. The total attenuation along the link is about 38 dB. Each 43 km fiber introduces 10 dB losses, and the 22 km round-trip between Université Paris 13 and Aubervilliers has a significant loss of 15 dB probably due to the large number of connectors and the OADMs. The optical input power is about 2 mW at Observatoire de Paris and the power injected into the 11 km fiber at Université Paris 13 is 35 µW.

The link-induced phase noise compensation system (Fig. 2) is based on our previous system described in [11]. The only change concerns the use of a bidirectional amplifier at the end of the link. At the remote end signal is separated into two parts, one is mixed with the laser signal to evaluate the transfer performance and the other part is sent back to the input end. At the input end, the link-induced phase noise is obtained from the beat-note between round-trip signal and the input signal. After filtering and amplification, the link phase noise is actively cancelled through a phase-lock loop using acoustooptic modulator as an actuator.

As a first result, the Internet traffic was unaffected during the whole period of the test (about 3 weeks). The bit error rate was continuously monitored and no error was detected. This was a crucial point since the University Paris 13 is the access point of a metropolitan area network of about 100 km serving hospitals, high schools and universities.

The link stability performance is evaluated from the end-to-end beat-note signal. The optical phase noise power spectral density of the 108 km link is shown in Fig. 3, without and with compensation. The link phase noise rolls down very sharply after a few hundred Hz. The phase noise reduction reaches around 65 dB at 1 Hz, which is coherent with the fact that, at low frequency, the rejection is limited by the link delay and scales as $\left( f \ t_{trip} \right)^2$ where $t_{trip} \approx 0.54$ ms is the 108 km trip time [9]. Note that the laser phase noise is sufficiently low to prevent any limitation to the performance of the stabilization [11]. Fig. 4 shows the fractional frequency Allan deviation of the 108 km link for four days of continuous operation (red circles). It was measured with a π-type dead-time free frequency counter from the end-to-end beat-note signal. Signal was filtered with a ~10 Hz bandwidth tracking filter. The free running fiber frequency noise is measured simultaneously from the frequency instability of the compensation signal (open circles). The Allan deviation is $4\times10^{-16}$ at 1 s averaging time and scales down with a $1/\tau$ slope from 1 s to 5000 s. After 10000 s of averaging time, it reaches a floor around $7\times10^{-20}$. Without filtering, the Allan deviation is 5 times higher between 1 s and 5000 s.

The comparison between the 86 km dedicated link and the present 108 km hybrid link allows insight into the phase noise contribution of the two 11 km fibers from the Internet network. According to Fig. 3, the lineic free running phase noise power spectral density of the 22 km RENATER fibers at 1 Hz (14 rad$^2$/Hz/km) is 10 times higher than that of the 86 km fiber. By the same time the delay-limited rejection should decrease proportionally to the square of the lengths ratio (2 dB). This is not clearly noticeable from Fig. 3 because the link noise rejection is limited by the delay only below 10 Hz or less [11]. With compensation, the integrated phase noise from 1 Hz to 1 kHz is increased from 0.4 to 1.5 rad (rms) for the 86 km and 108 km links respectively. Concerning the Allan deviation, the additional noise of the two 11 km fibers increases the 108 km link instability by a factor slightly below 2 between 1 and 40 s compared to the 86 km link (see Fig. 4). For longer time, both stabilities are limited by the noise floor of the compensation system. Note that the 86 km long-term stability was improved compared to previous results [11] due to a better stability of the measurement interferometer. However, comparison with other link noise measurements [10, 12-13] shows that the lineic phase noise of a dark fiber can span over three orders of magnitude (0.07.rad$^2$/Hz/km [13] to 180 rad$^2$/Hz/km [12] at 1 Hz), with a noise roll off law between $f^2$ and $f^{-3/2}$ in 1Hz-100 Hz region. Moreover, it varies slightly in time. Thus any consideration concerning the noise scaling with the link length is very difficult.

This 108 km link represents a first step towards long-distance frequency transfer using fiber network infrastructure carrying simultaneously internet traffic. For longer distances, an obvious solution consists in splitting the link into shorter compensated segments and developing intermediate stations. Each one should achieve three functions, to send back part of the received signal to the previous station, to amplify and filter the received signal ("repeater" function), and to compensate the phase noise induced by the next segment. The repeater could be a laser phase-locked to the incoming signal. However, use of a standard telecommunication network brings several difficulties. As already pointed out, Internet traffic is unidirectional, whereas our transfer system requires bidirectional propagation inside each fiber segment. Second, standard telecommunication networks use optical to electrical conversion of data signal in all nodes and regenerators, whereas RS dissemination has to be purely optical from end to end. Therefore most of the network equipment would have to be equipped by a specific bypass for the RS channel. Finally, there is highly restricted access to the fiber network infrastructure, and the remote intermediate stations have to work independently without human intervention.

We have demonstrated an ultra-stable optical link which uses partly an optical telecommunication network simultaneously carrying Internet data. Frequency transfer was demonstrated with instability of $4 \times 10^{-16}$ at 1 s and integrates down to $7 \times 10^{-20}$ after 10000 s. Such an optical link allows the transfer of present and near future optical clocks without any stability degradation. Further steps will be to take advantage of the RENATER 600 km fibers from Aubervilliers to the German border near Strasbourg, where cross-border fibers connect the two French and German scientific networks. The total attenuation of the various segments is 167 dB. Hence several intermediate stations should be implemented in order to regenerate the signal and increase the control bandwidth. The demonstration of such a transnational 600 km dual link is very challenging for the future development of continental-scale wide-range frequency transfer.

The authors are deeply grateful to D. Vandromme and F. Simon from GIP RENATER, and J. F. Florence from Université Paris 13, for their support in using the Internet network between University Paris 13 and Aubervilliers. We acknowledge funding support from the Agence Nationale de la Recherche (ANR BLAN06-3_144016).

over 534 km", in *Conference on Lasers and Electro-Optics/International Quantum Electronics Conference and Photonic Applications Systems Technologies*, Technical Digest (CD) (Optical Society of America, 2004), paper CTuH4.

**Figure captions**

Figure 1: Schematic of the optical frequency transfer over dedicated and non-dedicated fibers (MAN : Metropolitan Area Network, WAN : Wide Area Network, corr : correction, WDM : Wavelength Division Multiplexing, ITU : International Telecommunication Union)

Figure 2: Schematic of the link compensation

Figure 3: Phase noise power spectral density of the (a) free running 108 km, (b) compensated 108 km link, and, in dotted lines, of the (c) free running 86 km, (d) compensated 86 km link.

Figure 4: Fractional frequency instability of the 86 km (open squares) and 108 km (open circles) free running link and 86 km (squares) and 108 km (circles) compensated link measured with a 10 Hz filter, and noise floor of the compensation system (blue triangles), versus averaging time.

Figure 1

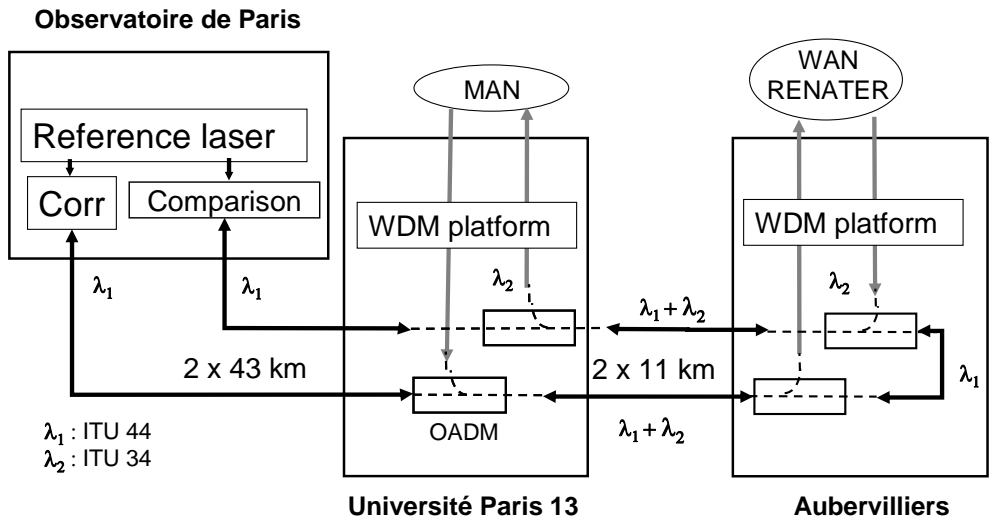

Figure 2

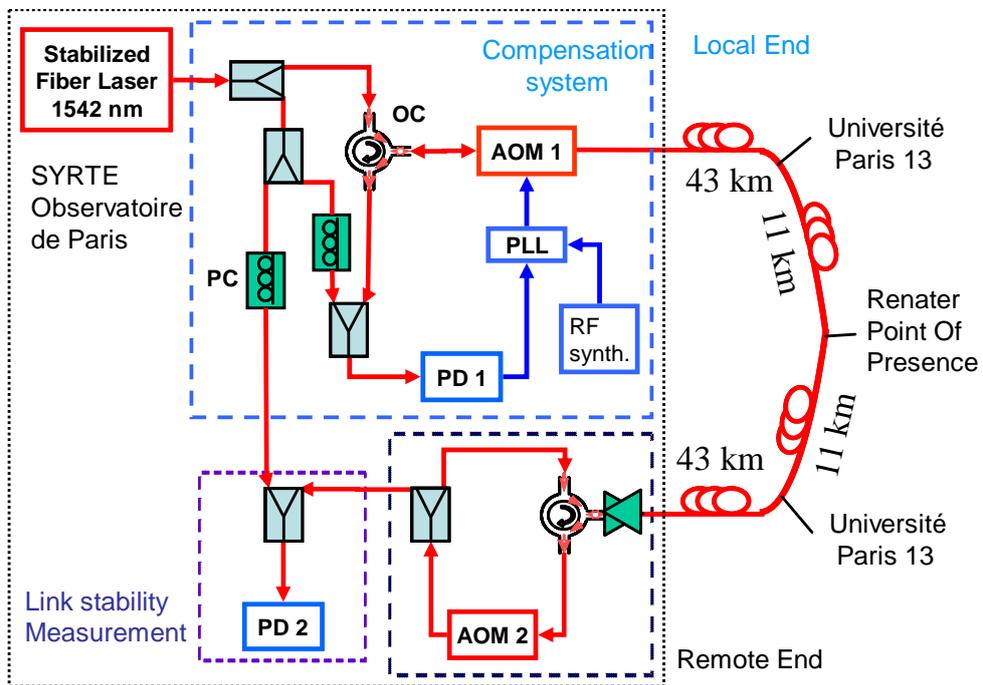

Figure 3

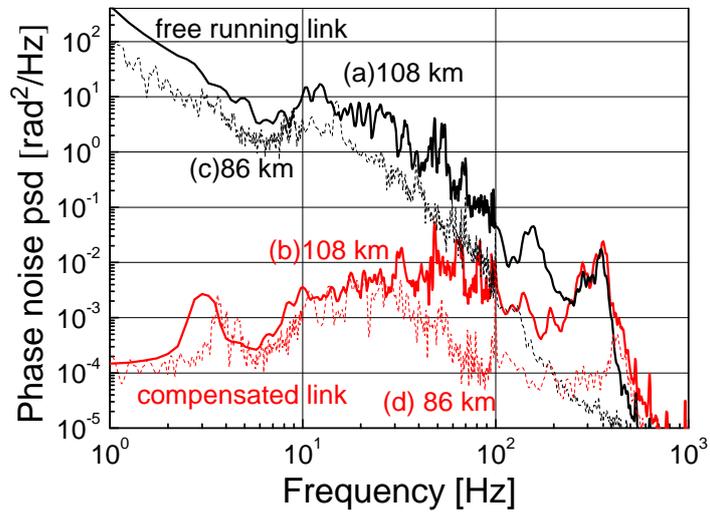

Figure 4

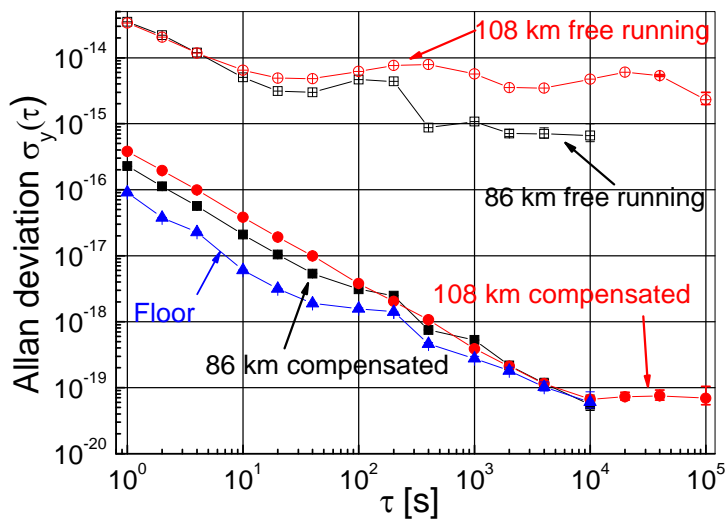